\documentclass[apjl]{emulateapj}
\usepackage{epsfig}

\begin{document}

\title{The Effect of Spiral Structure on
the Stellar Velocity Distribution in the Solar Neighborhood
}

\author{A.~C.\ Quillen}
\author{Ivan Minchev}

\affil{Department of Physics and Astronomy, University of Rochester, Rochester, NY 14627;
 aquillen@pas.rochester.edu,iminchev@pas.rochester.edu}

\begin{abstract}
Clumps in the solar neighborhood's stellar velocity distribution
could be caused by spiral density waves. 
In the solar neighborhood,
stellar velocities corresponding to orbits
that are nearly closed in the frame rotating with a spiral pattern
represent likely regions for stellar concentrations.
Via particle integration, we show that
orbits can intersect the solar neighborhood
when they are excited by Lindblad resonances with 
a spiral pattern. 
We find that a two-armed spiral density wave with pattern speed placing the 
Sun near the 4:1 Inner Lindblad Resonance (ILR)
can cause two families of nearly closed orbits in the solar neighborhood. 
One family corresponds to  square shaped orbits
aligned so their peaks  lie on top of, and support,
the two dominant stellar arms.  The second family correspond
to orbits $45^\circ$  out of phase with the other family.
Such a spiral density pattern could account for two major clumps in the solar
neighborhood's velocity distribution. 
The Pleiades/Hyades moving group corresponds to the first family of orbits 
and the Coma Berenices moving group corresponds to the second family.
This model requires a spiral pattern speed of approximately
$0.66 \pm 0.03$ times the angular rotation rate of the Sun or 
$18.1 \pm 0.8$ km s$^{-1}$ kpc$^{-1}$.
\end{abstract}


\section{Introduction}

The velocity distribution of stars in the solar neighborhood contains
structure which has been particularly clearly
revealed from recent studies of Hipparcos observations
\citep{dehnen98,skuljan,famaey,chereul98,chereul99,nordstrom04}.
Much of this structure was previously identified with moving
groups \citep{eggen}.  Moving groups are groups of
stars which are kinematically associated.
Young, early-type stars can be moving together because they carry the kinematic
signature of their birth \citep{eggen}.  However, a number of the kinematic
clumps identified through the kinematic studies also contain
later type and older stars \citep{dehnen98,famaey,nordstrom04}.
Dubbed superclusters \citep{famaey,eggen},
these kinematic associations of stars, which are observed all over
the sky, can share the same space motions as well-known open clusters.
The best documented of the superclusters are 
the Hyades supercluster associated with the Hyades cluster, 
the Sirius or Ursa Major
supercluster, associated with the Ursa Major cluster,  
and the Pleiades moving group (or Local Association) associated 
with young clusters such as the Pleiades.

There have been a number of studies exploring the origin of moving
groups and superclusters. 
\citet{eggen} and others 
have proposed that the structure in the velocity
distribution was a result of inhomogeneous star formation and the dissolution
of clumps of stars formed simultaneously.
Large moving groups could be produced by the dissolution of large stellar
agglomerations associated with spiral arms (e.g., \citealt{asiain99}).
However, the presence of older
stars in the moving groups or superclusters  has presented
a challenge for this simplest scenario.
To account for the older stars
\citet{chereul98} suggested that
superclusters could be superpositions of different age moving groups.
Because some of the moving groups contain stars spanning a range colors,
groups could be long lived and may not necessarily have transient features
associated with recently formed clusters that are being disrupted.
\citet{desimone}
suggested that the clumps in the velocity distribution 
are due to irregularities in the Galactic
potential.  In their model, spiral arms near corotation
were induced stochastically on a distribution of stellar test particles
causing localized structure in the stellar velocity distribution.

The possibility that we consider here is that the clumps in
phase space (in the velocity distribution) are associated with
or caused by spiral density waves traveling in the solar neighborhood,
as discussed by \citet{famaey,dehnen98,desimone}.
This type of explanation would not be unprecedented.
The Hercules stream at a tangential velocity of 
$v\sim 45$km/s has been explained in terms
of perturbations caused by the Galactic bar \citep{dehnen00,fux,raboud}.
This stream is due to stars in orbits strongly affected by the Galactic Bar's
2:1 Outer  Lindblad Resonance.
Based on this elegant explanation for the Hercules stream, we are
motivated to 
explore similar explanations for structure in the velocity 
distribution, but at velocities nearer those of 
circular orbits.
Before we consider the effect of spiral density
waves on the solar neighborhood's velocity distribution,
we first review what is known about the strength, number of
arms and possible pattern speed (angular rotation rate) for spiral density waves
likely to be propagating in the solar neighborhood.

The recent study of \citet{vallee02} provides a good summary of 
the many studies that have used observations to map the Milky Way disk.
Cepheid, HI, CO and far-infrared observations suggest that the Milky Way 
disk contains a four armed tightly wound structure, whereas \citet{drimmel}
have shown that the near-infrared observations are consistent with a dominant
two-armed structure.
The dominant two-armed and
weaker four-armed structure was previously proposed by \citet{amaral}.
The nearest spiral arm (excluding the local Orion armlet) is 
the Sagittarius/Carina arm, 0.9kpc away from the Sun 
in the direction toward the Galactic center.  The distance
between this arm and the Perseus arm in the opposite direction 
from us (toward the Galactic anticenter) is about 2.5kpc.

Based on observational constraints on the Galactic spiral structure,
a number of studies have created dynamical models to 
fit kinematic observations. 
These models are sensitive to, and so constrain, the spiral pattern speed. 
Reviewing previous work, \citet{shaviv} finds a clustering
of estimates for the pattern speed of local spiral structure near 
$\Omega_s \sim  20 {\rm  km s^{-1} kpc}^{-1}$, though other
studies suggest $\Omega_s \sim 13 {\rm  km s^{-1} kpc}^{-1}$. 
\citet{lepine} suggest that locally the Milky Way can be 
modeled by the superposition of a two- and four-armed structure.
Their model places the Sun near the corotation resonance 
$\Omega_s \sim  28 {\rm  km s^{-1} kpc}^{-1}$),
and was fit to Cepheid kinematics.
The recent gas dynamical studies \citep{martos, bissantz} 
match the properties of the gas in nearby arms
with a spiral pattern speed of $\sim 20 {\rm km s^{-1} kpc}^{-1}$.
\citet{martos} propose that a two-armed stellar structure consistent
with the stellar distribution inferred from COBE
could cause four-arms in the gas distribution near the Sun. 
The gas dynamical model proposed by \citet{bissantz} with a similar spiral
pattern speed matches HI and CO kinematics.

In this paper we consider the effect of a spiral density wave on
the solar neighborhood velocity distribution.   In section 2, we
discuss how the velocity components of stars
near the Sun are related to their orbits in the Galaxy.
In section 3 we describe our technique for determining 
the velocity components of different populations of stars.
In section 4 we explore how different spiral density waves
perturb the velocity distribution.  A summary and discussion follows.


\section{Epicyclic motion and the position on the $u,v$ plane}

We must relate the observed velocities of stars
to quantities describing their orbital motion in the Galaxy.
The solar neighborhood velocity distribution can be 
described as a function of the azimuthal velocity component, $v$,
and the radial velocity, $v_r=-u$, where $u>0$ corresponds
to velocities toward the Galactic center.  We define $v$ such that
the tangential component of the velocity, in the direction
of Galactic rotation, is equal to $V_0 + v$.
where $V_0 = 220 $km/s, 
is the local standard of rest at the position of the Sun 
for a Galactocentric radius of $R_0 = 8$kpc \citep{reid}.

In the absence of perturbations from
spiral arms, the motion of stars in the disk of a
galaxy can be described in terms of radial or epicyclic oscillations
about a circular orbit.  It is useful to specify
the relation between the observed
velocity components $u,v$ and parameters that describe
the epicyclic motion.   These parameter are the mean radius
or guiding radius $r_g$ and the epicyclic amplitude.
The energy of an orbit in the plane of an axisymmetric 
system (neglecting perturbations from spiral structure) is
\begin{equation}
E(u,v) = {(1 + v)^2\over 2}  + {u^2 \over 2} + \ln{r}
\label{uv}
\end{equation}
where the potential energy, $\ln{r}$, is that appropriate for a flat rotation
curve, and $r$ is the Galactocentric radius.
In the above equation we have put velocities in units of $V_0$
and radii in units of $R_0 = 8.0$ kpc, the distance
of the Sun from the Galactic Center.
The angular rotation rate of a star in a circular orbit
at the Sun's radius from the Galactic Center,
$\Omega_0 = 28 {\rm km~s}^{-1}\ {\rm kpc}^{-1}$ which is based
on observations of the proper motion
of Sag A*  \citep{reid} and is consistent with measurements based on
Tycho and Hipparcos observations \citep{olling}.
Henceforth we place angular rotation rates and pattern
speeds in units of $\Omega_0$.

In an epicyclic approximation  we can write the energy
\begin{equation}
E = {1\over 2} + \ln{r_g} + E_{epi}
\label{epi1}
\end{equation}
where ${1\over 2} + \ln{r_g}$ is the energy of a star in a circular orbit
about a guiding radius  $r_g$ and $E_{epi}$ is the energy from
the epicyclic motion.
\begin{equation}
E_{epi} = {u^2 \over 2} + {\kappa^2 (r-r_g)^2 \over 2}
= {\kappa^2 a^2 \over 2}
\label{epi2}
\end{equation}
where $a$ is the epicyclic amplitude and $\kappa$ is the epicyclic
frequency at the guiding or mean radius $r_g$.
We can also write $E_{epi} = j \kappa$ where $j$ is the
radial action variable.

We now consider stars specifically in the solar neighborhood
restricting us to a specific location in the Galaxy.
The velocity distribution in the solar neighborhood, 
or number density of stars as a function of $u,v$ is measured near the Sun
where $r\approx 1$ in units of $R_0$.   
Using $r=1$ in equation (\ref{uv}) and setting the energy
equal to that written in terms of the epicyclic motion
using equations (\ref{epi1}, \ref{epi2}), 
we solve for $r_g$.   It is convenient to define the distance between 
the guiding or mean radius and the Sun's Galactocentric radius, $s=r_g -1$.
To second order in $v$ we find that a star near the Sun with velocity components
$u,v$ has a guiding radius with 
\begin{equation}
s \approx v
\end{equation}
and epicyclic amplitude
\begin{equation}
a \approx \sqrt{{u^2\over 2} + {v^2}}.
\end{equation}

The above two relations allow us to relate the velocity components
$u,v$ for stars in the solar neighborhood,
to quantities used to describe the epicyclic motion; the guiding
radius and epicyclic amplitude. Particles with positive $v$ have $s>0$
and so guiding radii that are larger than $R_0$, and mean radii outside
the Sun's Galactocentric radius. Particles with negative $v$ have $s<0$
and so are expected to spend most of their orbits inside $R_0$.  The
distance from the origin $u=v=0$ determines the epicyclic amplitude.

The angular rotation rate of a star is most sensitive to the value of its 
guiding radius $r_g$.  The location of Lindblad resonances with 
a periodic perturbation are therefore primarily set by a star's guiding radius.  
In the solar neighborhood the guiding radius is primarily dependent on the 
$v$ velocity component.  The location of resonances on the $u,v$ plane
is therefore most sensitive to the $v$ value.
We focus on the location of resonances because they are one possible 
cause for the structure in the solar neighborhood velocity distribution.   
In the vicinity of a resonance there is a bifurcation 
in the families of periodic orbits 
and there are {\it no orbits} that have low
epicyclic motion (e.g., \citealt{cont75}).
Because resonances can induce large epicyclic (or radial)
motions, stars with guiding or mean radii distant from the solar 
neighborhood could be seen in our vicinity,
as is the case for the Hercules stream and the Galactic Bar's 
Outer Lindblad Resonance. 

\section{Finding orbits that are near periodic orbits}

In this section we describe our numerical procedure for 
constructing a function that can be compared to 
the number density of stars $f(u,v)du dv$
in the solar neighborhood. 
We are restricting this initial study to a spiral perturbation moving at a unique fixed pattern speed or angular rotation rate $\Omega_s$.
In this case the Jacobi integral is conserved. 
By transferring to the frame rotating with the pattern, the Hamiltonian can be 
brought to a time independent form (e.g., \citealt{B+T}).

We wish to weight positions in the $u,v$ plane at the position
of the Sun according to how likely they are to be occupied by stars. Since
stars are born in spiral arms with low velocity dispersion, we assume that
they are born in orbits that are nearly periodic or closed in the frame
rotating with the arms.  After a star is born, it can be scattered 
by molecular clouds or transient spiral structure as it
orbits in the Galaxy.  Older
stars would occupy orbits that oscillate at a greater
extent about the periodic
or closed orbits.  We have chosen to weight orbits 
according to how near they are to closed or periodic orbits.
Consequently we construct a numerical way to measure the extent
of oscillation about a periodic orbit for stars in the solar neighborhood.

For each position in the $u,v$ plane, we carry out a numerical 
integration of a particle with 
initial condition  $r=1, \phi=0$ 
corresponding to its location near the Sun.
Its initial radial velocity $v_r = -u$
and tangential velocity $1+v$ (in the inertial frame).  
In the frame rotating
with the spiral pattern, we integrate the trajectory of
the particle.  Each time the particle passes through $\phi=0$
(the Sun-Galactic center line), we record its position and velocity 
components.  After a specified number of orbits (in most cases 10), 
we compute the variance of radii ($\sigma_r$)
and $u$ values ($\sigma_u$) at times when the particle passed through $\phi=0$.
The weighting function we used to estimate the likelihood of populating the
position in the $u,v$ plane
is the sum of these two variances or $W=\sigma_r + \sigma_u$.
We adjusted the number of orbits integrated so that the resulting
weighting function was fairly stable; doubling the number of
orbits integrated did not result in large changes in $W$.
By keeping the number of integrated orbits low, we could
explore a larger region of parameter space.

We use our weighting function to identify locations on 
the $u,v$ plane likely to be populated with stars.
Orbits with initial $u,v$ values that have low values
of our weighting function $W$ correspond to orbits where a young star is likely
to be born.  Older stars could remain in these locations but also would
also be scattered out to larger values of $W$ corresponding
to locations in the $u,v$ plane with fewer stars.

Our procedure for constructing a weighting function 
is related to that used to construct surfaces of section.  
Surfaces of section can be computed by fixing the Jacobi integral 
and plotting $u$ vs $r$ each time the particle passes through $\phi=0$
(e.g., \citealt{B+T}).  
In a surface of section fixed points correspond to 
periodic or closed orbits in the galaxy.
In such a diagram, our weighting function describes the square of the 
distance in phase space from a fixed point.  
We do not use surfaces of section here because they are generated for 
each value of the Jacobi integral. We wish to study the likelihood of 
populating orbits as a function of velocity in the solar neighborhood
so we desire a weighting function that is a function of $u,v$.

Our method can be compared to the backwards integration technique used 
by \citet{dehnen00}.  Stars at each $u,v$ were integrated
backward in time while the Galactic bar was reduced in strength.  
Velocities on the $u,v$ plane corresponding to 
orbits with initially low epicyclic motions (before the
growth of the Bar) were given higher
weights \citep{dehnen00}.  
The theory of adiabatic invariants implies that particles
with initially low epicyclic motion settle onto 
periodic or closed orbits following the slow growth of a perturbation.  
Consequently, we expect that the backwards integration method produces 
weighting contours similar to those of our weighting function.
Compared to the backwards integration method 
our weighting technique  has the advantage 
that it is insensitive to 
the initial stellar velocity distribution and   
the manner of perturbation growth (see \citealt{fux}).
These assumptions are a particular problem for spiral density waves
since they could vary in both amplitude (as considered by \citealt{desimone})
and wavevector or pitch angle 
(as considered by \citealt{fuchs01a,fuchs01b}).

Our particle integration is done for particles moving 
in the plane of the Galaxy.  
We assume a gravitational potential 
$V(r, \phi, t) = V_0(r) + V_1(r,\phi,t)$ and 
$V_0(r) = \ln r$ corresponding to a flat rotation curve.
We assume that the gravitational potential perturbation $V_1(r, \phi, t)$ 
is caused by tightly wound logarithmic spiral density perturbations.  
Here $r$ and $\phi$ are the radius and azimuthal angle in the Galactic plane
and $t$ is time.  
The potential perturbation can be expanded in terms of Fourier components
\begin{equation}
V_1(r, \phi, t) = \sum_m A_m \cos(\alpha_m \ln r + m (\phi - \Omega_s t - \gamma_m)).
\end{equation}
If only one Fourier component is present, $m$ is the number of
arms of that component.  The angle $\gamma_m$ corresponds 
to an angular offset measured
at time $t=0$. 
Measured at the Galactic Center, $\gamma_m$ is
the angle between the location of the Sun ($r=1, \phi=0$) 
and the peak of the spiral pattern at the same radius 
but at an azimuthal angle differing from that of the Sun.  
A subsequent figure illustrates this angle.
The parameter
$\alpha_m$ depends on the pitch angle, $p_m$ where $\alpha_m = m \cot p_m$.
Note $\alpha_m < 0$ for trailing arms when the rotation is clockwise.  
Maps of the Milky Way are commonly shown from the view point of
an observer located above the Galaxy in the direction of the North
Galactic Pole (e.g., \citealt{drimmel,vallee02}.  
On these maps the rotation is clockwise.

In the WKB or tight winding approximation, the amplitude of the potential 
perturbation Fourier component is related to the density perturbations 
\begin{equation}
A_m \sim { - 2 \pi G  \Sigma_0 S_m R_0 \over |\alpha_m| V_0^2}
\end{equation}
(e.g., \citealt{B+T}).
In the above equation  we have placed $A_m$ in units of $V_0^2$.
Here $\Sigma_m$ is the amplitude of the $m-$th  Fourier component
of the spiral mass surface density and $S_m = \Sigma_m/\Sigma_0$.
The mean surface density of disk mass in the solar neighborhood
is $\Sigma_0 \sim 50 M_\odot {\rm pc}^{-2}$ \citep{holmberg}.
We estimate for the solar neighborhood
\begin{equation}
A_m \sim 0.03  S_m 
  \left({ \Sigma_0 \over 50 M_\odot {\rm pc}^{-2}}\right)
  \left({ 220 {\rm km s}^{-1} \over V_0^2 }\right)^2
  \left({7 \over \alpha_m }\right)
  \left({R_0 \over 8 {\rm kpc}}\right).
  \label{Am}
\end{equation}
A two-armed spiral is seen in the near infrared COBE/DIRBE 
data with pitch angle in the range  $p \sim 15.5-19^{\circ}$,
corresponding to  $\alpha = m \cot p$ in the range 5.8-7.2 for
$m=2$ \citep{drimmel}. 

We now estimate ballpark values for the potential
perturbation component $A_2$.
Using $S_2 \sim 0.15$ consistent with parameters
described by \citet{drimmel} for the stellar component 
(based on K band observations)
the above equation \ref{Am} gives us $A_2 \sim 0.005$. 
Specifically \citet{drimmel} estimated
$(\Sigma_{max} - \Sigma_{min})/\Sigma_{min}=0.32$, though they
suggested that the true value could be higher.

\section{Structure in the $u,v$ plane }

\begin{figure*}
\epsscale{0.3}
\plotone{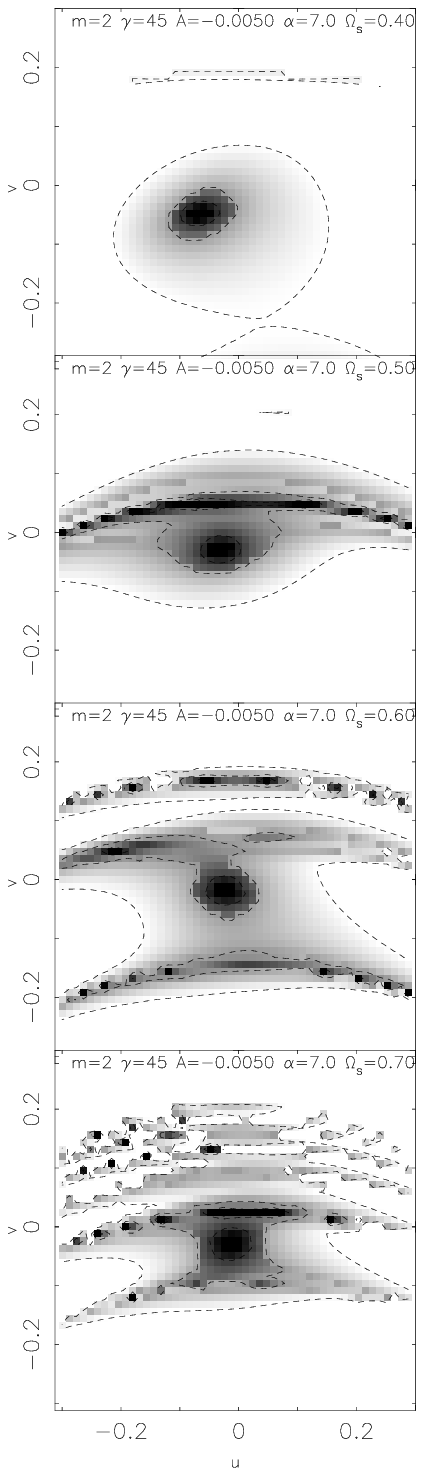}
\figcaption{
The $(u,v)$ plane shown for two-armed spiral models at different pattern speeds.
The $u,v$ axes are shown in units of $V_0$, the velocity of a star
in a circular orbit at the Sun's Galactocentric radius. 
The gray scale and contours
show our weighting function $W$ which gives a measure of the 
extent of epicylic motion about a closed or periodic orbit.  From top 
panel to bottom panel we show
the effect of spiral structure with pattern speed $\Omega_s = 0.4, 0.5, 0.6$  and $0.7$
respectively (in units of $V_0/R_0$).  The two armed perturbations have  angular
offset $\gamma_2 = 45^\circ$, wavevector with $\alpha = 7$, and potential
perturbation strength $A_2=0.005$ (see Table \ref{table:all}).
Dark regions correspond to orbits which are nearly periodic.  
The top panel is nearest
the 2:1 ILR and so closed or periodic orbits (in the frame of
the pattern) are highly elliptical.  This causes the periodic orbits
to be moderately distant from the origin where $u=v=0$ and where circular
orbits are located in the absence of spiral perturbations.
The bottom panel is near the outside the 4:1 Lindblad resonance.  
The structure at low $v$ in the bottom panel
is caused by the 4:1 Lindblad resonance.
At larger $v$ structure from higher $m$  Lindblad resonances is encountered
closer to corotation.  
\label{fig:twoarmed}
}
\end{figure*}

Because the Milky Way has a dominant two-armed spiral pattern,
we consider dominant two-armed spiral density waves.
In Figure \ref{fig:twoarmed} we show the structure of our weighting function
$W(u,v)$ for a pure cosine two-armed 
spiral pattern with parameters listed in Table \ref{table:all}.
Hereafter Table \ref{table:all} lists parameters used
to describe the spiral perturbation.  Each line in the table
is labeled with a corresponding figure displaying 
the output of a simulation.

The strength of the spiral pattern of the simulation shown
in Fig \ref{fig:twoarmed}
is near but somewhat above that estimated by \cite{drimmel} with
$A_2 = -0.005$. 
Peaks in density correspond to dips in the gravitational potential.    
We use negative Fourier amplitudes
so that $\gamma_m$ refers to the location
of the peaks (with respect to the Sun)
of the spiral density pattern. The density
perturbation has the opposite sign of the potential perturbation.

Each panel in Fig \ref{fig:twoarmed} 
corresponds to spiral structure moving at a different pattern speed.
We now relate positions in the $u,v$ plane to those expected for
Lindblad resonances.
The $m:1$ Inner Lindblad resonance is located where
\begin{equation}
\Omega_s = \Omega ( 1 - \sqrt{2}/m).
\end{equation}
for a flat rotation curve, where $\kappa=\sqrt{2}\Omega$
and $\Omega$ and $\kappa$ are computed as a function of  the guiding radius.
Here we adopt the notation $m:1$ ILR referring to $m$ epicyclic oscillations
for each orbit around the galaxy in the frame moving with the pattern.
Specifically,
for the 4:1 ILR $\Omega_s = 0.65 \Omega$, 
for the 3:1 ILR $\Omega_s = 0.53 \Omega$, and 
for the 2:1 ILR $\Omega_s = 0.29 \Omega$.
The topmost panel of Figure \ref{fig:twoarmed}
with $\Omega_s = 0.4$ (in units of $\Omega_0$)
has the Sun outside but near the 2:1 ILR. 
The bottom  panel (with $\Omega_s = 0.7$)
has the Sun just outside the 4:1 ILR and so nearer to corotation.

In the images shown in Figure \ref{fig:twoarmed} 
the log of our weighting function, 
$\log_{10} W(u,v)$ is  shown in gray scale.
Contours are shown with the highest contour with $\log_{10} W = -1$ and 
a contour interval of 1. 
The blackest regions correspond to smallest values of $W$ with
$\log_{10} W \sim -4$ which are those most likely to be populated
by young stars. These regions could also maintain older 
stars at low velocity dispersion.

In Figure \ref{fig:twoarmed} structure is primarily influenced
by the value of $v$, as expected from our discussion in section 2;
since $v$ sets the location of resonances with the spiral pattern.
For larger values of $v$ the angular rotation rate about the Galaxy drops.
As $v$ varies, resonances with the spiral perturbation are crossed.
Consequently orbits affected by resonances that have
large epicyclic excursions can be seen 
in the Solar neighborhood.

Dark regions on these diagrams correspond to nearly closed orbits
in the frame moving with the spiral pattern.
The strongest concentration of closed or periodic orbits in all
panels shown in Figure \ref{fig:twoarmed}
is near the origin where $u=v=0$ and orbits
are nearly circular.  When the spiral pattern
nears the 2:1 ILR (top
panel with $\Omega_s =0.4$), the periodic orbits become elongated
(highly elliptical) and so their position in the $u,v$ plane is
shifted away from the origin.    The orbits are elliptical
and so have non-circular velocity components.
On the lower part of this panel 
stars in orbits perpendicular to those supporting the
arm can be seen.  
Stars on these orbits have guiding radii on the other side of the 2:1 ILR.

\begin{figure*}
\epsscale{1.0}
\plotone{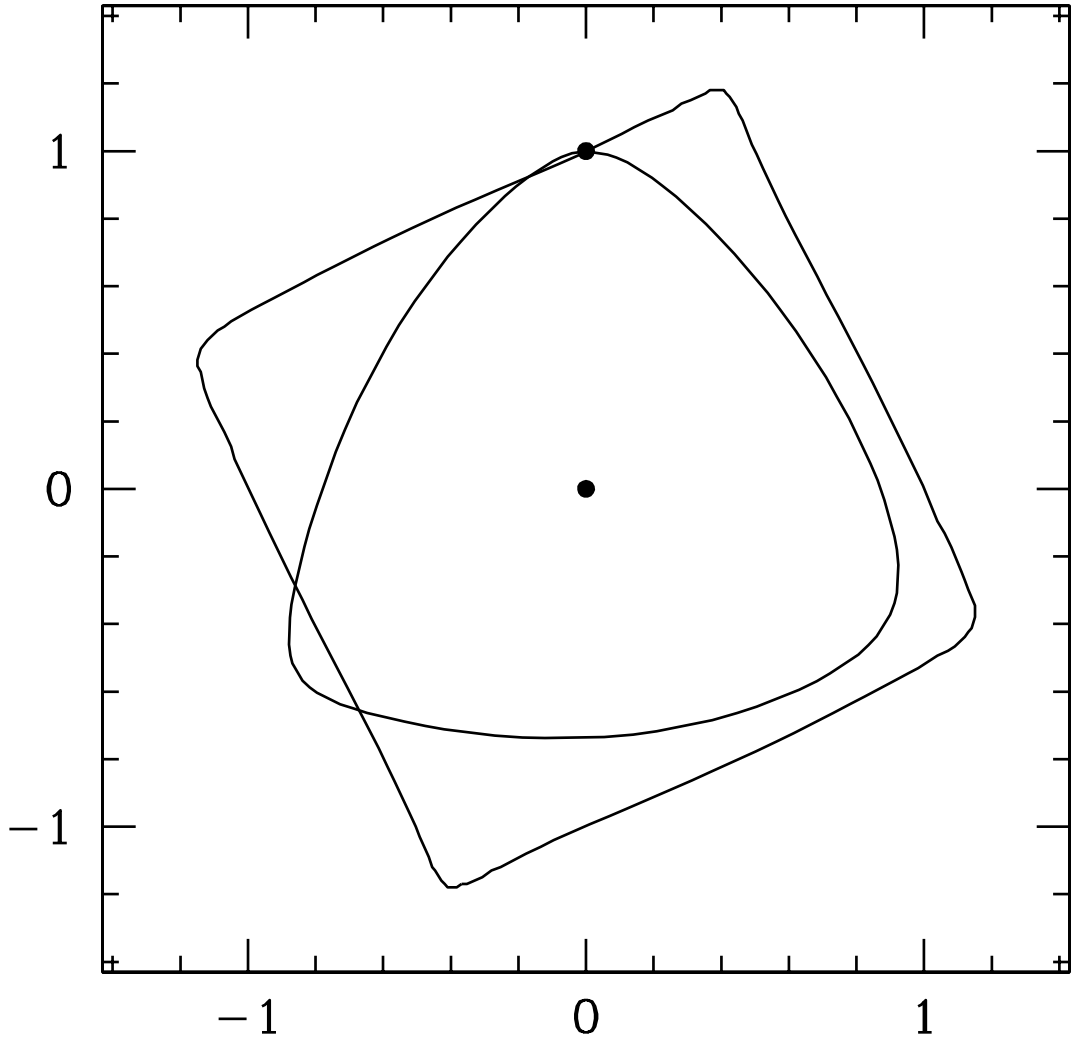}
\figcaption{
Closed orbits in the frame moving with the
spiral structure are shown 
for the two-armed $\Omega_s = 0.60$ model (see Table \ref{table:all} for parameters). 
The weighting function for this model is shown as 
the second panel from the bottom
in Figure \ref{fig:twoarmed}.
Orbits are shown for two dark regions shown on this panel.
The triangular closed orbit corresponds to position
on the $u,v$ plane with $u=0.00$, $v=-0.15$.
The square orbit corresponds to $u=-0.20$, $v=0.05$.
\label{fig:Uorb}
}
\end{figure*}

In these panels features are also seen at locations separate
from the largest concentration at the origin.
In the second panel from the bottom, structure is seen
near $v\sim 0.05$.  In this simulation the Sun is just inside
the 4:1 ILR. These orbits correspond to square shaped orbits, 
elongated because of this resonance.
To illustrate these we show accompanying orbits for
these locations in the $u,v$ plane.
In Figure  \ref{fig:Uorb} orbits associated with two of the
dark regions on this panel are shown.
The structure at $v\sim 0.15$ in the panel on figure \ref{fig:twoarmed}
corresponds to triangular shaped orbits influenced by 
the 3:1 ILR.  This orbit is also shown in Figure \ref{fig:Uorb}.

Nearer to corotation (the lowest panel in \ref{fig:twoarmed}
at $\Omega_s = 0.7$) orbits above and
below the central family are associated with orbits on either side
of the 4:1 resonance (supporting and opposite to the pattern).
At higher $v$, higher $m$ Lindblad resonances
are encountered closer to the corotation resonance.
Here many resonances influence the weighting function and so
the model stellar velocity distribution.

\subsection{Near the 4:1 Inner Lindblad resonance}

In the previous section we discussed two-armed patterns 
over a large range of pattern speeds, but at only one angular
offset, $\gamma_2 = 45^\circ$.  
We noted in Figure \ref{fig:twoarmed} 
that at higher pattern speeds more structure
appears in the weighting function on the $u,v$ plane as higher 
$m$-Lindblad Resonances are encountered closer to corotation.
In this section we consider in more detail
the possible role of the 4:1 Inner Lindblad resonance.
Because square shaped orbits are excited near this resonance we
expect that the velocity distribution could
depend strongly on the orientation of the spiral pattern. 
Therefore we must
consider the sensitivity as a function of the angular offset $\gamma_2$.

\begin{figure*}
\plotone{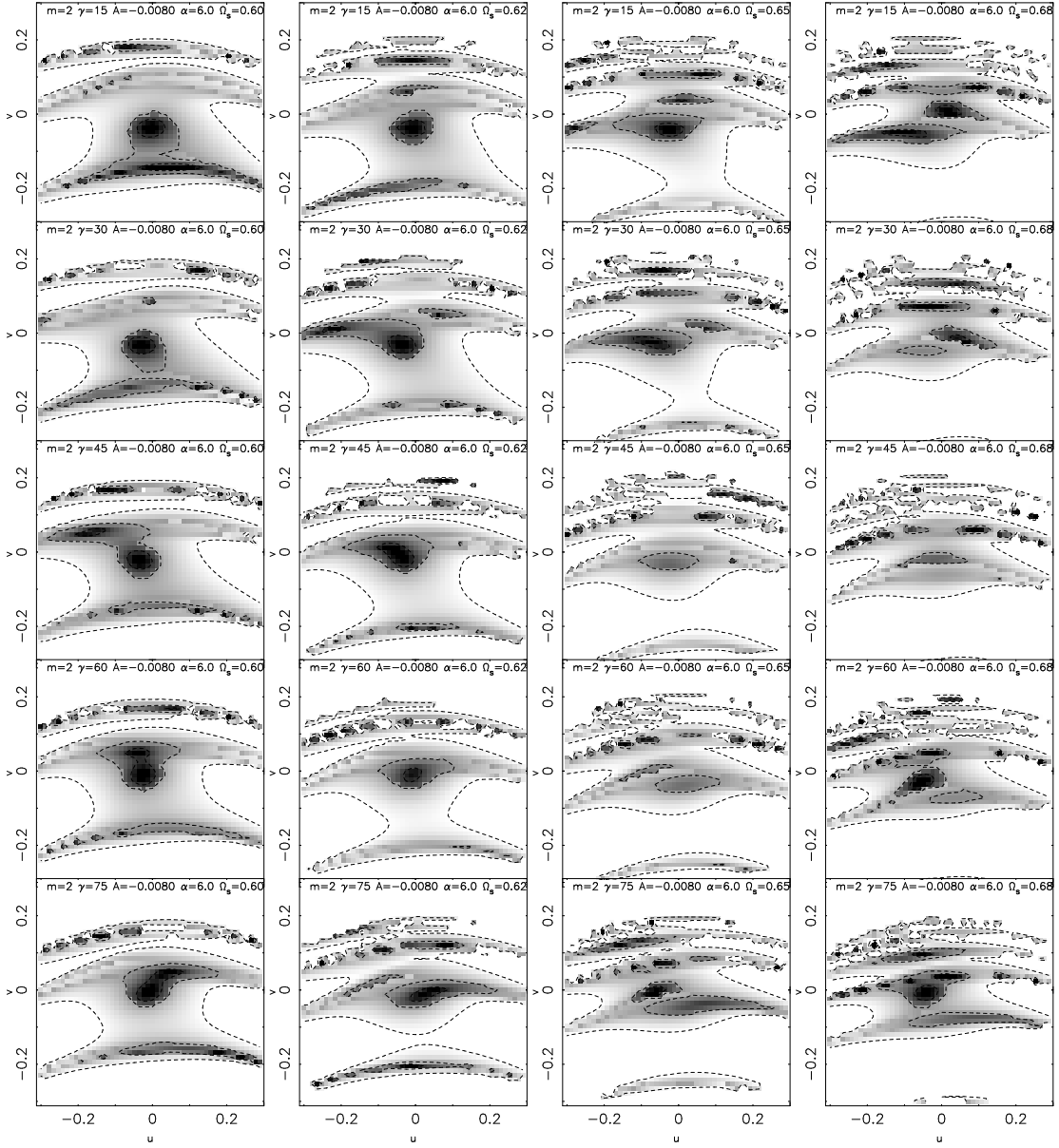}
\figcaption{
The $(u,v)$ plane shown for stronger two-armed spiral models 
at pattern speeds placing the Sun near the 4:1 Inner Lindblad resonance.
Panels are similar to those shown in Figure \ref{fig:twoarmed}.
From the leftmost column to the rightmost the pattern speed is 
$\Omega_s =0.60, 0.625, 0.65$ and 0.675 respectively.
In each column from top to bottom the angle of the spiral
pattern with respect to the Sun is $\gamma_s = 15, 30, 45, 60$ and $75^\circ$,
respectively. Parameters for these simulations are also listed
in Table \ref{table:all}.
\label{fig:strong}
}
\end{figure*}

In Figure \ref{fig:strong} weighting on the $u,v$ plane
is shown for a somewhat stronger two-armed spiral pattern.
Parameters for the spiral pattern are listed in 
Table \ref{table:all}.
From top to bottom on these panels, the angular
offset between the spiral maximum and the Sun 
$\gamma_2$ = 15, 30, 45, 60 and $75^\circ$.
From left to right on this figure the spiral
pattern speed $\Omega_s = 0.60, 0.625, 0.65$, and 0.675.
In Figure \ref{fig:strong} we note that a wealth of structure
exists in these plots particularly for the pattern speeds near and
above the 4:1 ILR at $\Omega_s\approx 0.65$.

\begin{figure*}
\plotone{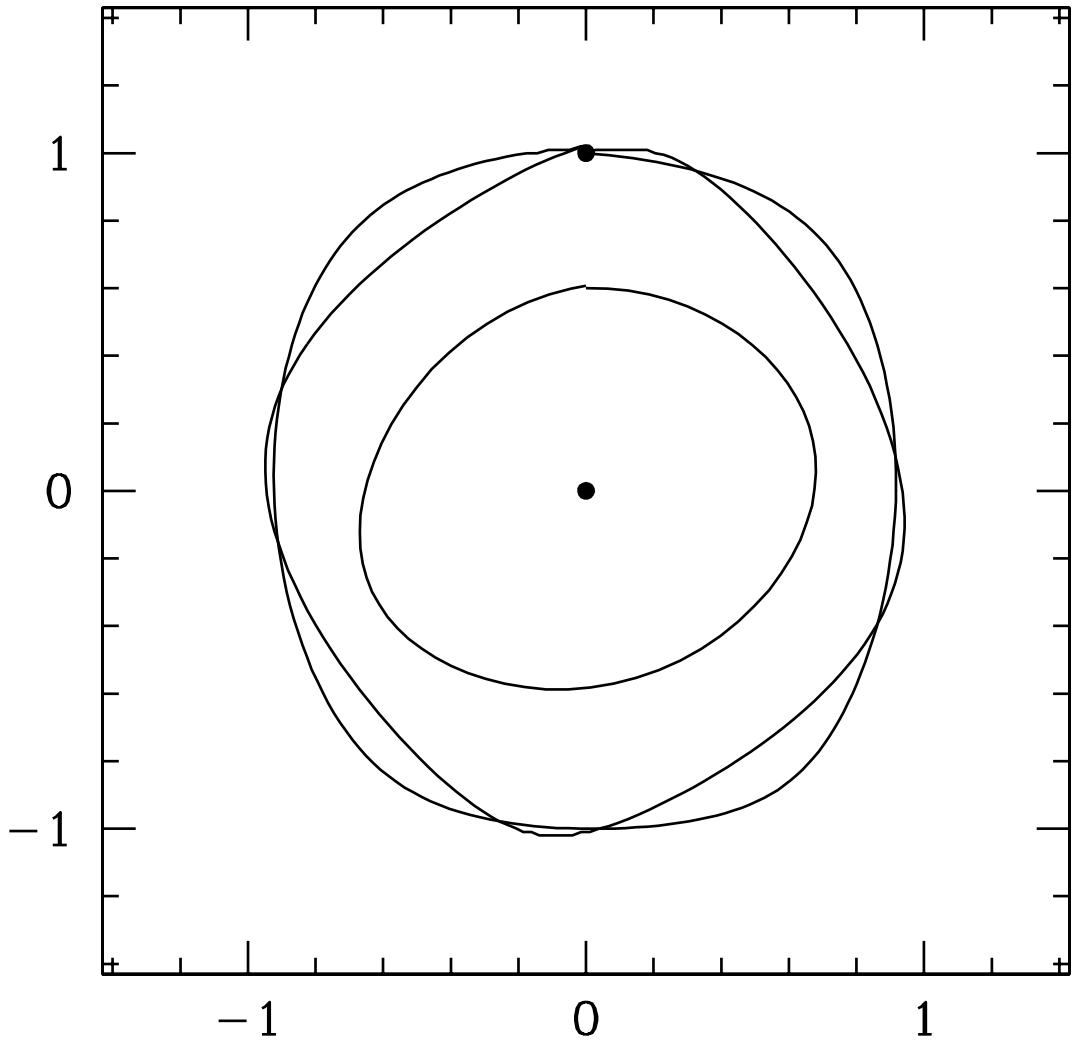}
\figcaption{
Closed orbits in the frame moving with the
spiral structure are shown 
for the two-armed $\Omega_s = 0.675$ model 
(see Table \ref{table:all} for parameters). 
The weighting function for this model is shown as 
the top panel in the third column from left
in Figure \ref{fig:strong}.
Orbits are shown for two dark regions shown on this panel 
at the location of the Sun.
The inner square closed orbit corresponds to position
on the $u,v$ plane with $u=-0.065$, $v=-0.043$.
This orbit supports (lies on top of) the dominant two-armed
perturbation.  In other words, two of the orbit peaks are on top
of two of the arms.
The outer rounder square orbit corresponds to $u=0.025$, $v=0.015$
and is out of phase with the two-armed density perturbation.
We also show an inner orbit supporting the two-arms inside the 
Galactocentric radius of the Sun.
\label{fig:Zorb}
}
\end{figure*}

If the spiral pattern places the Sun near its 4:1 ILR,
we see that
two dominant clumps in the weighting function appear near the
origin in the $u,v$ plane for low values of $\gamma_2$ 
(top panels on rightmost two columns in Figure \ref{fig:strong}).
In Figure \ref{fig:Zorb} we show the shape of closed orbits 
(in the frame rotating with the pattern) for these two
clumps for $\gamma_2 = 15^\circ$ and $\Omega_s = 0.675$ (also
see Table \ref{table:all}). 
Orbits are plotted for $(u,v)= (-0.065,-0.043)$, the inner diamond orbit
and $(u,v) = (0.025,0.015)$, the outer rectangular orbit.
The inner orbits would have density peaks at the location
of two of the density peaks aligned with the density perturbation.
Consequently, we say they support the spiral pattern.
The outer orbits have guiding radius on the other
side of the 4:1 ILR and so are out of phase with
the inner orbits. 

As pointed out by previous works, two-armed density perturbations
can excite square shaped orbits 
\citep{cont86} and a four armed gaseous response \citep{martos} near the 4:1 ILR.
At guiding radii exterior to the 4:1 ILR, a pure $m=4$ potential 
perturbation (with potential components $A_4 \neq 0$, $A_2 = 0$),  
excites orbits that support the spiral pattern,
Interior to the 4:1 ILR, the closed orbits would be out of phase 
with the density perturbation and
so would fail to support the spiral pattern.
The orientation or phase of the orbits changes at the resonance.
When the orbits support the spiral pattern, they lie on top
of the perturbation in such a way that a self consistent
model can be created. In other words the density perturbation causes
orbit perturbations which in turn are consistent with the density
perturbation.
When there is only a strong two-armed potential perturbation, 
as is the case currently discussed here,
the orbits can still be strongly influenced by the 4:1 ILR.
In this case the 4:1 ILR is second order in the epicyclic amplitude ($\propto j$
rather than $j^{1/2}$; \citealt{cont86}). 
Again the closed orbits can be
square shaped near the 4:1 ILR, but are 
oriented so that they support the spiral pattern {\it interior} to the 4:1ILR
and fail to support the pattern {\it exterior} to the 4:1 ILR.
The tendency for orbits to fail to support spiral structure
outside the 4:1 ILR prompted \citet{cont85} to propose
that two-armed spiral patterns end near their 4:1 ILRs (also see 
\citealt{cont86,patsis}).

We now compare the structure seen in Figure \ref{fig:strong}
with the velocity components of moving groups.
In our units dominant moving groups in the solar neighborhood
velocity distribution have  velocity components
$(u,v) = (-0.05  , -0.10)$ for the Pleiades moving group,
        $(-0.18  , -0.09)$ for the Hyades moving group,
        $(-0.04  , -0.02)$ for the Coma Berenices moving group, and
        $( 0.04  ,  0.01)$ for the Sirius/Ursa Major moving group.
These velocity components are based on the
stellar distribution measured with Hipparcos observations and
are given with respect to 
the Sun by \citet{dehnen98}.  The study by \citet{famaey} grouped
the Pleiades and Hyades moving groups together and 
identified the Coma Berenices and Sirius/UMA groups as separate structures.
Note we have divided the velocities by $V_0$.

Based on the rightmost two columns in Figure \ref{fig:strong}
which are at pattern speeds $\Omega_s =0.65, 0.675$ near the 4:1 IRL,
it is attractive to associate the Pleiades/Hyades moving group
with the nearly closed orbits
at $v\sim -0.05$ and the Coma Berenices moving group with
the nearly closed orbits  near the origin,
($u\sim v\sim 0$).
For $\gamma_2 \sim 15^\circ$, the $u,v$ weighting function shows
an extension to low $u$ for the lower clump, 
suggesting that low oscillation orbits
exist over an elliptical region in the $u,v$ plane that
could encompass both the Hyades and Pleiades moving groups.
If this choice of spiral perturbation accounts for
the Hyades/Pleiades and Coma Berenices moving groups
then the Hyades/Pleiades moving group stars are in orbits
oscillating about the interior diamond shaped closed orbit passing
through the solar neighborhood (see Figure \ref{fig:Zorb})
and the Coma Berenices group stars are near the outer rectangular 
shaped closed orbit.
We note that
both families of closed orbits are elongated or not perfectly 
square. This would be expected because
the potential perturbation is elliptical or two-armed. 
The square shape of these closed orbits is a result
of the proximity of the 4:1 ILR. 

\begin{figure*}
\plotone{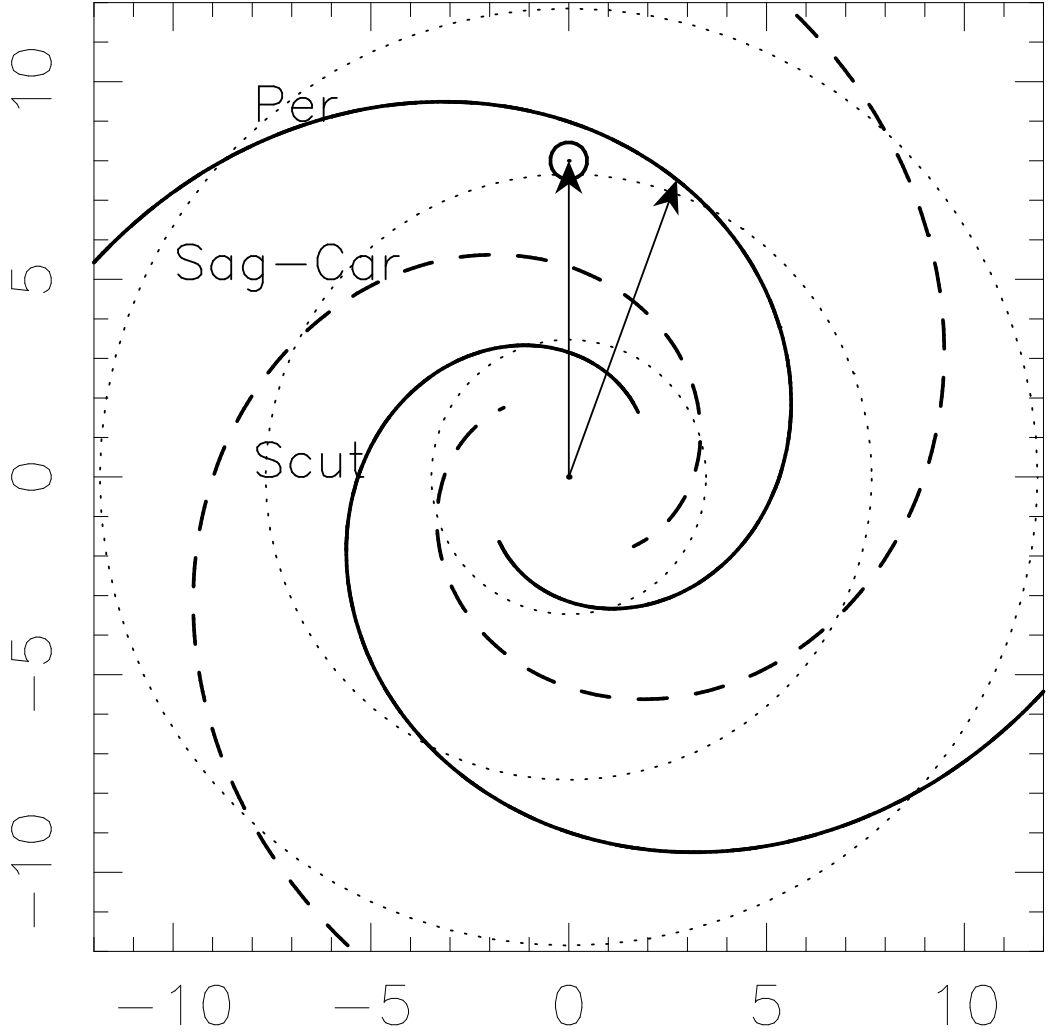}
\figcaption{
Location of arms with respect to the Sun consistent 
with the distributions shown in Figure \ref{fig:strong} with
$\gamma_2=15^\circ$ (top panels).  
The two stellar arms are shown as solid lines.
If two additional arms were present in between the two  strong
stellar ones, they would be located approximately 
at the dashed lines.
The position of the Sun is shown as a small circle.  The location of the
2:1 ILR, 4:1 ILR and corotation resonance are shown as large dotted circles.
The angle between the Sun and nearest strong arm, $\gamma_2$,
is that between the two vectors.  
We have labeled the arms according to their common names.
In this figure Galactic rotation is clockwise.
See Table \ref{table:all} for descriptions of the parameters.
\label{fig:arms}
}
\end{figure*}

We have found that
the 4:1 ILR provides a promising explanation for some of
the structure in the solar neighborhood velocity distribution.
If a spiral density wave can produce this structure, then it
should be consistent with the location of spiral arms
near the Sun.
We now consider the location of the spiral arms for a model
that could account for the Hyades/Pleiades and Coma Berenices moving
groups.  Of the panels shown in Figure \ref{fig:strong}
we choose that with $\Omega_s=0.675$ and $\gamma_2 = 15^\circ$.
The position of the two spiral arm density peaks 
consistent with this model are shown as solid
lines in Figure \ref{fig:arms}.  If there were an additional two arms
located in between the dominant two stellar ones, they would
be located where the dotted lines are shown on Figure \ref{fig:arms}.
Axes on this figure are given in kpc so that
the spiral arms in the model can be compared with 
the location of observed arms (based on names referred to 
by \citealt{vallee02}).

From Figure \ref{fig:arms} we
see that our spiral model successfully places the Sun
between the Perseus and Sagittarius/Carina arms.
We have adjusted the pitch angle of the spiral arms (set $\alpha$) 
so that the separation between the Perseus arm and
the Sagittarius/Carina arm is consistent with that suggested
by \citet{vallee02}.  However, our model requires that the Sun is
closer to the Perseus arm than the likely location of
the Sagittarius/Carina arm,
in contradiction to the model by \citet{vallee02}.
We note that the Sagittarius/Carina arm is not as
strong in stars as it is in gas \citet{drimmel}, and 
the dust map of \citet{drimmel} shows a kink or bend
in the Sagittarius/Carina near the Sun.  This bend
may be associated with the change in the orbit structure
at the 4:1ILR.  The model of \citet{vallee02} assume purely
a logarithmic shape for the spiral arms, whereas the
orbit structure near the 4:1 probably causes
deviations in the location of the arms. 
Fourier decomposition of nearby galaxy images has shown
that even when the galaxy is grand design,
Fourier components with $m>2$ may still be significant
(e.g., \citealt{elmegreen,seigar,grosbol}) and the simple model
explored here only considers a dominant $m=2$ Fourier component.
A more detailed model for the arms (based on the observations) and 
a self consistent model for the spiral arms,
the gas response and population of orbits would
be required to determine if the arm locations were consistent
with the proposed dynamical model.

We now discuss sensitivity of our proposed spiral wave model to
the parameters used to define it.  We have carried out
comparison simulations with different
spiral arm pitch angles.  We find that the weighting function contours 
(and inferred structure in the velocity distribution)
are not strongly sensitive to the value of $\alpha_2$.  
For $\alpha_2$ between 5-7,
consistent with the range of pitch angles discussed by 
\citealt{drimmel,vallee02},
little difference is seen the structure of
the weighting function.
However, from Figure \ref{fig:strong} it is clear
that the model velocity distribution (and orbital structure)
is very sensitive to the orientation angle of the spiral arms ($\gamma_2$)
with respect to the Sun.
Changes in $\gamma_2$ of as small as $10^\circ$ can cause significant changes in
the locations of closed orbits  in the solar neighborhood.

We have also varied the strength of the two-armed perturbation, $A_2$.
For $|A_2| < 0.005$ the orbits are more nearly circular and
would not be able to be consistent with two dominant spiral arms (the orbits
are not sufficiently elliptical away from the 2:1 ILR).
For $|A_2| < 0.005$ the width of the 4:1 ILR is smaller (the resonance
is second order the epicyclic amplitude) and the two families
of orbits near the 4:1 ILR are closer together, too close
to account for the Hyades/Pleiades and Coma Berenices moving group
separations.   For $|A_2| > 0.01$ the separation between
the two orbit families was larger than that of the two moving
groups. We had best results finding a model $u,v$ distribution
similar to that observed in the solar neighborhood 
for $A_2  \sim 0.008 \pm 0.002$.
We note that this is about twice as high as that
estimated from the K-band model by \citet{drimmel}.
Specifically \citet{drimmel} estimated
$(\Sigma_{max} - \Sigma_{min})/\Sigma_{min}=0.32$ which
is smaller than the contrast seen in
other similar nearby galaxies \citep{seigar,grosbol}. 
Consequently \citet{drimmel} 
commented that they suspected the true value for
the density contrast (arm vs interarm) could be higher 
than this number. 

\begin{figure*}
\epsscale{1.0}
\plotone{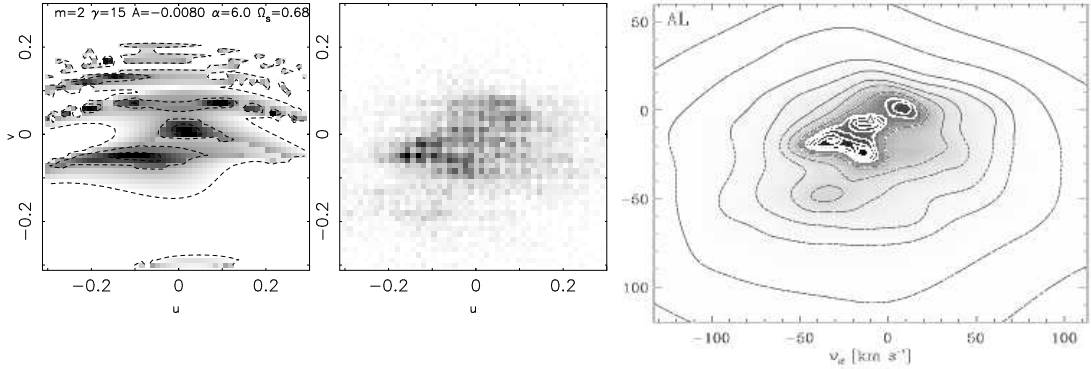}
\figcaption{
Comparison between our model for the $u,v$ plane (leftmost
panel) and the observed stellar velocity distribution.  
The left hand panel (similar to Figure \ref{fig:twoarmed}) has
parameters listed in Table \ref{table:all}.
The middle panel is a 2D histogram of the $u,v$ velocities
of the F and G dwarf stars listed by \citet{nordstrom04} in units
of $V_0$.  The $u,v$ velocity components listed in Table 1
by \citet{nordstrom04} are heliocentric.  To compare these velocities to
those shown in the lefthand panel we have assumed a VLSR velocity
of $U_0\approx V_0 \approx  8$km/s.
The right hand panel is from \citet{dehnen98} Figure 3 for all
stars in his Hipparcos subsample.
In the right panel, the clump nearest
the origin is the Coma Berenices moving group.  The clump
below the Coma Berenices group is the Pleiades moving group.
The clump at positive velocities is the Sirius/UMA moving
group and the clump at $v=-40$km/s is the Hyades moving group.
We note that our model predicts that the Hyades and Pleiades
moving groups are kinematically related.  This is consistent with
the study by \citep{famaey} which found no strong
separation between them.  Our model is consistent with the
velocity separation
between the Coma Berenices and Hyades/Pleiades moving groups
and predicts a slight tilt (larger $v$ for more negative $u$)
to the Hyades/Pleiades moving
groups contours that is seen in the observed velocity distribution.
In our model, the Hyades/Pleiades stars are on diamond shaped
orbits supporting the spiral pattern and those in the Coma Berenices
group are on orbits that are out of phase by $45^\circ$. 
The Hyades/Pleiades stars have mean radii that are within 
the Sun's Galactocentric radius.  Those at larger $v$ have
mean radii outside $R_0$.
\label{fig:compd}
}
\end{figure*}

In Figure \ref{fig:compd} we directly compare our model
(see Table \ref{table:all} for parameters)
$u,v$ plane panel
showing our weighting function, to the observed stellar 
velocity distribution in the solar neighborhood.  
The middle panel on this figure is the stellar velocity distribution
of the F and G dwarf stars listed by \citet{nordstrom04}.
The right hand panel of this figure is
from Figure 3 of  \citet{dehnen98} 
which shows all stars in his Hipparcos sample. 
This panel from Figure 3 by \citet{dehnen98} resembles 
Figure 20 by \citet{nordstrom04} and Figure 7 from \citet{famaey}.
In our panel the origin corresponds to a circular orbit in the
absence of perturbations by spiral arms, however  the stellar
velocity distribution is plotted with respect to the solar
velocity.  If our proposed dynamical model is correct
it suggests that the Sun is moving with $u,v> 0$.

In the Figure by \citet{dehnen98} the clump nearest
the origin is the Coma Berenices moving group.  The clump
below the Coma Berenices group is the Pleiades moving group.
The clump at positive velocities is the Sirius/UMA moving
group and the clump at $v=-40$km/s is the Hyades moving group.
We note that our model predicts that the Hyades and Pleiades
moving groups are kinematically related.  This is consistent with
the studies by \citet{famaey,nordstrom04} which found no strong
separation between them.  Our model is consistent with the 
velocity separation
between the Coma Berenices and Hyades/Pleiades moving groups
and predicts a slight tilt (larger $v$ for more negative $u$)
to the Hyades/Pleiades moving
group contours that is seen in the observed velocity distribution.

Based on the comparison between our model and
the observed velocity distribution, it is tempting to associate the
Sirius/UMA moving group with closed orbits at higher $v$
that would correspond to higher $m$ ILRs.  However since
these would be affected by higher $m$ resonances, the structure of the orbits
is likely to be very strongly dependent on the location
of the Sun with respect to the spiral structure and on the
assumed structure of the spiral arms. Consequently we don't
feel we can attach any significant to structure in
our model at the location of the Sirius/UMA moving group.
If higher order resonances affect stars with guiding radius
just outside the Sun, then it's possible that the outer  Milky
Way disk is flocculent \citep{flocculent}.
It is also possible that an additional spiral density wave
could affect the outer part of the galaxy, adding an additional
complication that we have not explored here.

The 4:1 Lindblad resonance provides an explanation for some of the structure
present in the stellar velocity distribution in the solar neighborhood.
However, this explanation succeeds only over a very narrow range of pattern
speeds.  
For the small range of reasonable perturbation strengths,
based on the structure seen in Figure \ref{fig:strong},  we estimate that
the spiral pattern speed must be within $\Omega_s = 0.66 \pm 0.03$.
This corresponds to a pattern speed of $\Omega_s = 18.1 \pm 0.8$ 
km s$^{-1}$kpc$^{-1}$ (after restoring the physical units).
This pattern speed is consistent with many previous
estimates \citep{shaviv} and the kinematic models
by \citet{martos,bissantz}.   Our model for the
velocity distribution is quite similar to that
of \citet{martos} who showed that a dominant two-armed stellar
perturbation could cause four arms in the gas distribution.


\section{Summary and Discussion}

In this paper we have considered the effect of spiral density waves
on structure in the stellar velocity distribution in the solar neighborhood.  
We find that Lindblad resonances
with spiral density waves can cause structure in the velocity distribution.
Lindblad resonances can excite large epicyclic perturbations, allowing stars
from distant locations  to reach the solar neighborhood.
In the solar neighborhood,
we find that the location of a resonance is primarily set by
the tangential velocity component, $v$. 
Because spiral perturbations are not strong, their widths 
are narrow and they only excite large epicylic
oscillations over a narrow range of 
guiding or mean radii. However because these epicylic oscillations can
cause stars to cross into the solar neighborhood,  they can still
cause structure in the solar neighborhood velocity distribution.

Because Lindblad resonances can cause significant changes in the
structure of orbits,
they provide a promising explanations for structure seen
in the solar neighborhood velocity distribution, or
moving groups and superclusters.
To explore this possibility we have searched
for orbits that are near closed or periodic orbits in the 
frame moving with the spiral pattern.
We have constructed a weighting function that 
estimates the distance in phase space of an orbit from
a closed or periodic orbit.
When a moderately strong two-armed perturbation is present 
with pattern speed at $0.66 \pm 0.03$ times
the angular rotation rate of the Sun,
placing the Sun  near the 4:1 Lindblad resonance, 
two regions near the origin in the $u,v$ plane exhibit
nearly closed orbits.  
These islands are likely to be populated
by stars spanning a range of ages, as are moving groups.   

For a two-armed model with pattern speed placing the Sun near
the 4:1 Lindblad resonance, and an angular offset of $\gamma_2 =15^\circ$,
the structure of our weighting function resembles the observed velocity
distribution.
Nearly closed orbits are located near existing moving groups.
The region at at negative $v$ in the $u,v$
plane is elongated toward negative $u$ and so could encompass
both the Hyades and Pleiades moving groups.
The region at positive $v$, can be associated with the Coma Berenices
moving group.
This model provides a possible explanation for some of the structure
observed in the solar neighborhood's velocity distribution.
The model succeeds only over a very narrow range of pattern
speeds; with 
$\Omega_s = 18.1 \pm 0.8$ km s$^{-1}$ kpc$^{-1}$ providing
a tight constraint on the angular rotation rate
of the spiral pattern. Our model is most similar
to and consistent with the model proposed by \citet{martos} in which
two dominant stellar arms excite a four-armed
gaseous response.

We have shown here that 
resonances with spiral patterns provide a promising way to explain 
clumps in the solar neighborhood velocity distribution.
In this paper we have only considered the response of stars 
to spiral density perturbations
but did not consider the response of the gas. 
In our model we have not constructed a stellar orbit distribution
consistent with the assumed density perturbation.
Future work should 
strive to create models that are more self-consistent.
Here we have not explored the sensitivity of the stellar distribution
with stellar birth site or age. 
However  \citet{dehnen98, famaey}
found that structure in the velocity
distribution depends on the type of star or stellar population. 
Here we used a crude weighting function to 
find nearly periodic orbits in the $u,v$ plane,  
however we have not produced model stellar density distributions.
Future work could consider the birth of stars in a disk supporting
a spiral pattern and explore ways to predict the location and
number of stars as a function of birth site and age.
Future work could also consider the sensitivity of the present day
velocity distribution to the way that spiral structure evolves.
For example it may be possible to differentiate between 
pure amplitude growth (as considered by 
\citealt{desimone}) from shearing density wave
models for which both amplitude and wavevector vary simultaneously 
(e.g., as considered by \citealt{fuchs01a,fuchs01b}).

There are some interesting consequences of our model.  The 4:1 resonance
is likely to cause deviations from a pure logarithmic spiral pattern
near the Sun.  Better models for the Galaxy could include
arms that deviate from logarithmic spirals.  Because the Sun may
be located near a Lindblad resonance causing large epicyclic amplitudes
in the stars, measurements of the velocity of the local standard of rest (VLSR) 
and Oort's constants
are likely to be biased.  By understanding and correcting for these biases,
measurements of these astronomical quantities may be improved.
In this paper we have not proposed a dynamical explanation 
for the Sirius/UMA moving group.  This group could be related
to a higher $m$ Lindblad resonance, or perhaps there is another
spiral density wave moving at a slower pattern speed at larger
Galactocentric radius.   Here we have suggested that the Hyades/Pleiades
moving groups are on orbits with mean radii within the Sun's Galactocentric
radius, and the Coma Berenices and UMA/Sirius moving groups are
on orbits with mean or guiding radii outside $R_0$.  It would be
interesting to see if these stellar populations have different
age and metallicity distributions.
Here we have not discussed 
the orbits of stars at velocities in between the moving
groups. Because
of the 4:1 ILR, these stars should have large epicylic amplitudes and
so could be part of a different stellar population than the 
moving groups.   

Here we have not considered the role of more than one spiral
density wave or the role of the Galactic Bar.
Stellar orbits that are not affected by a Lindblad resonance 
from an additional perturbation, such as from
another spiral density wave or the Galactic Bar, would
be only weakly affected by the additional perturbation.
Away from a resonance, the orbital kinematics can
be treated with low order perturbation theory (for example as done 
by \citealt{B+T}). 
Only stars (at $v \sim 50$ km/s) associated with the Hercules stream 
that are influenced by the Bar's 2:1 Lindblad resonance,
should be strongly affected by the Galactic Bar. 
Stars at lower values of $|u|,|v|$, 
should be distant and so unaffected by Lindblad resonances 
with the Galactic Bar.
We expect that the closed orbits considered in this paper weakly oscillate 
or are weakly perturbed
at the frequency of the Galactic Bar. These
oscillations should cause small variations in
the location of the periodic orbits on the $u,v$ plane as 
seen from the Solar neighborhood.
However, as long as we consider orbits that are not associated with
resonances with the Galactic bar, then strong features associated
with the Bar would not be seen in the velocity distribution.
The same situation is likely if there is an additional spiral density
wave (at a different pattern speed) present in the solar neighborhood 
(as suggested most recently by \citealt{naoz}).
We only expect strong structures in the velocity distribution 
that are associated
with resonances from one of the spiral density waves present
in the Solar neighborhood. 
However, the presence of more than one perturbation can influence
the stellar dynamics.
For example, if the Solar neighborhood is
affected by more than one spiral density wave, then stars
at velocities between moving groups
might be on chaotic orbits \citep{galchaos}, a factor 
which could cause a relatively large increase in their 
velocity dispersion with time.

\acknowledgements
This work could not have been carried out without helpful
discussions with Larry Helfer and Don Garnett.
We thank the referee, B. Fuchs, for helpful comments
which have improved the manuscript.
%
%
%
A.~C.~Q.~gratefully thanks the Technion for hospitality and support 
during the fall of 2001 where this project was initiated.



{}

\clearpage

\vfill\eject

\clearpage

\begin{deluxetable}{lcccc}
\tablewidth{0pt}
\tablecaption{Parameters describing spiral patterns\label{table:all}}
\tablehead{
\colhead{Figures} &
\colhead{$A_2$}   &
\colhead{$\gamma_2$}      &
\colhead{$\alpha_2$}      &
\colhead{$\Omega_s$}     
}
\startdata
\ref{fig:twoarmed}  
        & -0.005  &  45$^\circ$ & -7     &  0.4,0.5,0.6,0.7   \\
\ref{fig:Uorb}  
        & -0.005  &  45$^\circ$ & -7     &  0.6   \\
\ref{fig:strong}      
        & -0.008  & 15,30,45,60,75$^\circ$ & -6  & 0.60,0.625,0.65,0.675 \\
\ref{fig:Zorb},\ref{fig:arms},\ref{fig:compd}  
        & -0.008  & 15$^\circ$  & -6    &   0.675
\enddata
\tablecomments{
The parameters of the spiral pattern corresponding to simulations
shown in the Figures.
The angular offset 
$\gamma_2$ in degrees are the angle (measured at the Galactic center)
between the location of the Sun and the nearest density maximum
of the spiral potential perturbation at a radius of $R_0$ 
(see Figure \ref{fig:arms}).
The perturbation strength $A_2$ is given in units of $V_0^2$ the velocity
of a star in a circular orbit at $R_0$.
The pattern speed, $\Omega_s$, is  given in units of $\Omega_0 = V_0/R_0$.
The parameter $\alpha_2$ sets the pitch angle of the spiral arms.
For $\alpha_2 <0$ the arms are trailing when the rotation is clockwise.
}
\end{deluxetable}

\end{document}